# INTERESTINGNESS – A UNIFYING PARADIGM
## *Bipolar Function Composition*

**Iaakov Exman**[1,2]

*Software Engineering Dept.[1],*
*The Jerusalem College of Engineering,*
*POB 3566, Jerusalem, 91035, Israel*
*iaakov@jce.ac.il*

*School of Engineering[2],*
*Bar-Ilan University,*
*Ramat-Gan, 52900, Israel*
*exmani@eng.biu.ac.il*

[Affiliation when original paper published in 2009]

Abstract:     Interestingness is an important criterion by which we judge knowledge discovery. But, interestingness has escaped all attempts to capture its intuitive meaning into a concise and comprehensive form. A unifying paradigm is formulated by function composition. We claim that composition is bipolar – i.e. composition of exactly two functions – whose two semantic poles are relevance and unexpectedness. The paradigm generality is demonstrated by case studies of new interestingness functions, examples of known functions that fit the framework, and counter-examples for which the paradigm points out to the lacking pole.

## 1   INTRODUCTION

Interestingness is an important criterion by which we judge discoveries, in particular knowledge discovery. But, interestingness has eluded all attempts to capture its intuitive meaning into a widely accepted formal framework.
    There are many proposals for the meaning of interestingness. Most of them fit our intuition to a greater or lesser extent. Some of them even correctly express one aspect or another of what should be interestingness. Though, none has convincingly covered the whole issue in a fundamental way.

This work starts with concepts firmly based upon our intuition to reach a unifying paradigm for interestingness. It is stated in terms of mathematical composition of exactly two functions, no less and no more. It enables grouping apparently disparate empirical formulas into a common paradigm.

Once its formal framework is made explicit, one can use it as a guide to propose functions to calculate interestingness and integrate them into novel knowledge discovery protocols.

A first example of interestingness composition involves a *matching* coefficient of a result as the relevance to an interest field, multiplied by the unexpectednes, given by a *mismatch* coefficient of the result to the same interest field. It simultaneously optimizes relevance and novelty of each result.

Moreover, the formal framework serves to check whether existing functions fit the unifying paradigm, or it points out to some required kind of addition.

An existing criterion to rank search data is given by the *TfIdf* formula. It perfectly fits the interestingness paradigm, as our analysis clarifies.

For each of the functions presented, we provide case studies, in a Web search setting, to demonstrate that they actually produce interesting results.

The remainder of the paper presents the unifying paradigm (section 2), introduces match-&-mismatch as an interestingness function pair (section 3), offers low-&-high-threshold computational functions (section 4), shows *TfIdf* in the new paradigm light (section 5), and ends with a discussion.

## 2   INTERESTINGNESS: THE UNIFYING PARADIGM

Interestingness, within knowledge discovery, is not an absolute quantity. It is variable along time and always relative to a field of interest. Here it is argued that it is bipolar, combining exactly two functions in a unifying paradigm for interestingness.

### 2.1   Exactly Two Functions: Relevance and Unexpectedness

Knowledge discovery means that we acquire new knowledge that we did not have previously. We use the term *Unexpectedness*, rather than the more neutral novelty, to emphasize that what is new is not strictly contained in any sense in the previously known.

The very meaning of Unexpectedness, as stated before, implies that it is relative to previously existing knowledge. We use the term *Relevance*, rather than relativity, to stress that the particular frame of reference is a chosen field of interest, e.g. material properties of metals, migratory birds, or software design patterns.

These two concepts, Unexpectedness and Relevance, are not just two faces of the same coin, one relative to the other, but two really independent functions. Indeed, they are separately quantified.

The previous knowledge exists whether or not new knowledge is acquired. Thus, the chosen field of interest can be characterized – say, by some kind of metadata – before and independently of any knowledge discovery event. In particular, the Relevance of any piece of knowledge, say a search result – be it novel or not – can be quantified relative to the reference metadata.

New knowledge is not determined by the previously known. In fact, one can acquire two pieces of new knowledge, having quite distinct contents, thus differently quantified, The Unexpectedness of one of them could be larger than the other one's, even with the same previous knowledge.

The first sense of interestingness is *Relevance* to the field of interest. In this sense, an item is interesting because it fits the field for which one has a rather stable interest, either professional or amateur. For instance, metals conduct heat, conduct electricity and have a shiny appearance. Copper concerns people interested in metals because it fits the context metadata.

The second sense of interestingness is *Unexpectedness*. In this second sense, an item is interesting because it calls one's attention by marked deviation from the typical item in the context. For instance, mercury is the only metal which is liquid at room temperature.

But, mercury is really interesting when one comes across it, because it is *both* a typical metal – conducting and shiny – and has unexpected properties – a liquid forming spherical drops.

Thus, we really have two functions, viz. Relevance and Unexpectedness.

Interestingness – A Unifying Paradigm                                      Iaakov Exman

## 2.2   No More Than Two Functions

We now argue that there are no more than two functions related to interestingness.

Suppose that we acquire a new piece of knowledge that by genetic modification, an agricultural station has developed yellow tomatoes, instead of the usual red ones. Yellowness certainly is a function that can be quantified by colorimetric methods and by accepted standards.

Should we include yellowness – or for that matter any other intrinsic property of the new knowledge – in the calculation of its interestingness? The answer is negative.

Let us look at the time dependence of interestingness, after such a knowledge discovery. At the discovery time, that piece of knowledge has a certain quantifiable amount of Unexpectedness.

The typical action after knowledge discovery is to incorporate the new piece into the body of knowledge available in the respective field of interest. The Unexpectedness of that piece of knowledge decreases dramatically. Interestingness decreases accordingly.

The situation is totally different with yellowness. A short time afterwards, the yellow tomatoes are still yellow. The time dependence of yellowness has nothing to do with one's knowledge of it.

Obtaining again the same piece of knowledge – a short time interval afterwards – will not be considered a discovery anymore. This is analogous to the registration of a new patent. Once a patent is registered, its contents are not novel anymore and it cannot be registered again.

Interestingness – through its Unexpectedness – is not an intrinsic property of any knowledge piece. It is a function of the discovery process.

## 2.3   The Formal Framework

The unifying paradigm for interestingness is formally expressed by the following equation:

$$\textbf{Interestingness} = \textbf{R} \circ \textbf{U} \qquad (2\text{-}1)$$

where **R** is a *Relevance* function,, **U** is an *Unexpectedness* function, and **o** is the symbol for function composition, meanimg either mathematical or computational function composition.

In equation (2-1) one should first apply the Unexpectedness function on the search results. On the respective output, one then applies the Relevance function, in this order. In its most general form, function composition is not commutative, in analogy to matrix multiplication.

One could conceivably think of other pairs of functions for which rather **U o R** is the appropriate order. But as long as the operator between functions is **o** the generic function composition, whatever is the order it cannot be reversed.

In certain particular cases, composition may be just multiplicative, and therefore commutative:

$$\textbf{U} * \textbf{R} = \textbf{R} * \textbf{U} \qquad (2\text{-}2)$$

Particular cases of importance are selection functions, that select items from the search result itemset, by specified criteria.

## 3   MATCH & MISMATCH

In our first example, a Match coefficient has the role of a *Relevance* function and a Mismatch coefficient is the *Unexpectedness* function. They are multiplied as in eq. (2-2) to obtain Interestingness values.

The case study illustrates these functions within a weblog recommendation system.

### 3.1   Match & Mismatch Coefficients

The Match coefficient is calculated by comparing the similarity between the field of interest context metadata and the current item.

The simplest Match coefficient is a boolean variable. It has value 1 if at least one keyword is common to both the field context and the current item. Otherwise it has a zero value.

A slightly more sophisticated Match coefficient uses an integer variable. Here one counts the common words of the context and current item.

A more complex Match coefficient calculation would involve in some way the edges in the metadata graph (say an ontology), where the vertices are keywords.

A Mismatch coefficient counts the dissimilarity between the field context metadata and the current item. It is an integer variable.

Assume the field metadata is a set F of keywords and the current item is a set C. In more precise terms, the mismatch is calculated as the symmetric difference of these sets:

$$\mathbf{Mismatch = F \triangle C = (F - C) \cup (C - F)} \qquad (3\text{-}1)$$

Note that mismatch terms can also be weighted, either by known keyword statistical factors or by the edges of an ontology graph.

Finally the expression for the Interestingness is:

$$\mathbf{Interestingness = Match * Mismatch / NormF} \qquad (3\text{-}2)$$

where NormF is a normalization factor that compensates for differences in the total number of keywords along result items.

### 3.2   Case Study: Weblog Recommendation

This is a case study of weblog recommendation, e.g. (Fujimura, 2005 [2]), (Glance, 2004 [3]). The user chooses a field of interest, here "sports". A context metadata is obtained from pre-defined data and user input.

Search was performed with the Yahoo search engine with the keywords: "Euro 2008 results and fixtures blog". Search results were post-processed to calculate values of Interestingness in equation (3-2). The sorted outcome is plotted in Figure 3.1.

The outcome is satisfactory for two reasons. First, there are few "interesting" results, so that one can make recommendations with confidence. These are seen in the graph of Figure 3.1. It clearly highlights only two results out of 93. One could set a threshold for automatic decision making.

Second, the Interestingness ranking is significantly different from the engine search ranking, which is based on very different considerations from interestingness. For instance, the two highest results by interestingness are ranked by Yahoo search in positions 29 and 67 respectively.
.

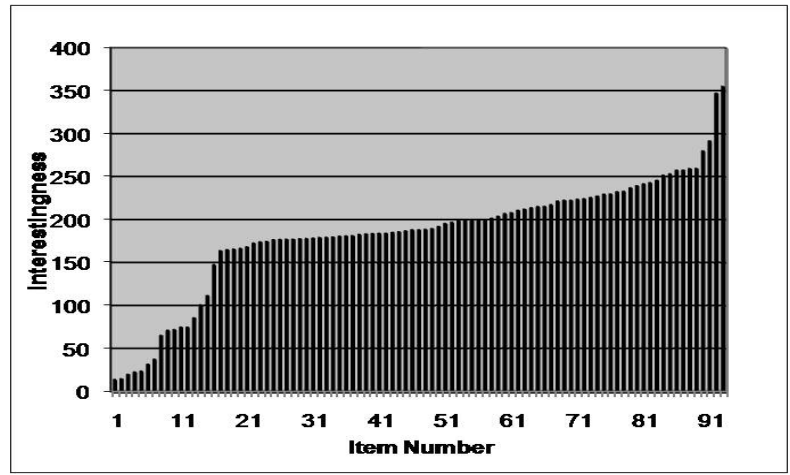

Figure 3.1: Interestingness plotted against item numbers for "sports" weblog items. Exactly two items have sharply higher values (around 350) than others.

## 4   LOW & HIGH THRESHOLDS

Our second example of interestingness calculation involves composition of computational functions as in eq. (2-1). This composition is not commutative.
The low threshold appears within the Unexpectedness function, while the high threshold is related to the Relevance function.

### 4.1   Low & High Computational Thresholds

The context here is discovery of new keywords. One first performs regular search in a field of interest for the user, using a standard search engine.
From the search result item-set one extracts all the non-trivial words and sorts them by frequency of appearance. Trivial words are articles, pronouns, propositions, etc., found in a "stopwords" file.
The *Unexpectedness* function **U** outputs all words below a low-frequency threshold **lowT**.
These candidate words are then tested by the *Relevance* function **R** as follows: repeat regular search with the same original keywords and each candidate word. The candidate is a new keyword if it is now above a high-frequency threshold **highT**.
Thus, the interestingness expression in this example has the same form as equation (2-1):

$$\textbf{Interestingness} = \textbf{R(highT) o U(lowT)} \qquad (4\text{-}1)$$

## 4.2	Case Study: Keyword Discovery

This case study refers to keyword discovery techniques, see e.g. (Arimura, 2000 [1]), (Moukas, 1997 [7]).

Samples of result sets with a size of 100 items were obtained from Yahoo web search of the keyword combination "migratory birds water swim".

The Unexpectedness function produced among the numerous low-frequency words appearing only once the word "phalarope" – previously unknown to us, as we are not ornithologists.

The Relevance function was applied next, with the candidate keyword "phalarope" added as input to the same keyword combination. The outcome shows this new keyword clearly above the higher threshold. This is seen in Figure 4.1, for results starting at result index $i$=1, up to $i$=100.

As counter-examples, the same threshold check for most other low-frequency words does not succeed.

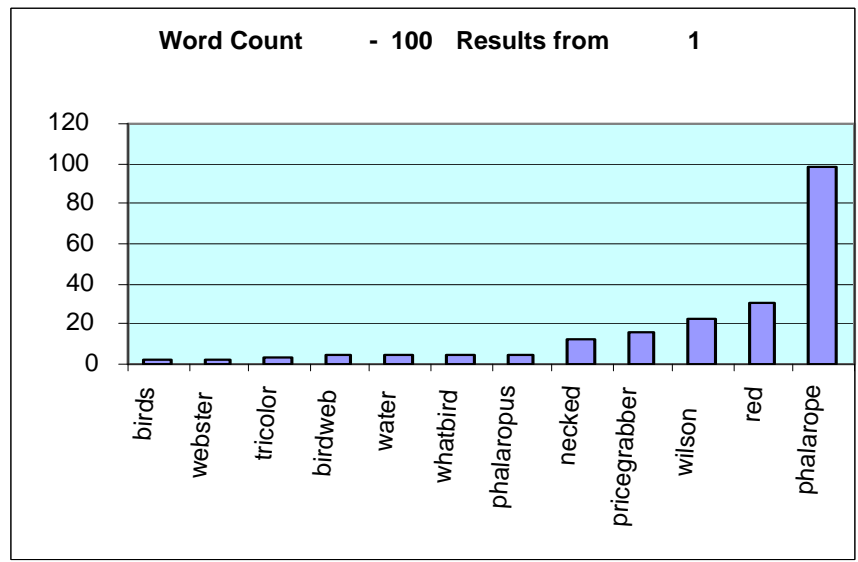

Figure 4.1: Relevance function highlights phalarope – In this histogram of selected words appearance, the candidate word is very prominent. Indeed it even appears in two forms (phalarope, phalaropus).

## 5	FREQUENCY & INVERSE FREQUENCY

**TfIdf** is a well-known ranking criterion for documents, given keywords of relevance – see e.g. (Yuwono, 1995 [11]). Here it is shown that it perfectly fits the generic framework for interestingness.

Its application is illustrated in reciprocal mode: to discover keywords of relevance, when given interesting documents.

## 5.1   Tf for Relevance

The so-called *Term Frequency* **Tf$_{jk}$** of keyword **k** in document **j** is:

$$\mathbf{Tf_{jk} = n_{jk} / T_j} \qquad (5\text{-}1)$$

where **n$_{jk}$** is the number of appearances of **k** in document **j**, and **T$_j$**, the total number of appearances of all terms **t** in document **j**, serves as a normalization factor.

**Tf$_{jk}$** is directly proportional to the number of keyword **k** appearances. When **k** is high-frequency in documents – as it is typical of keywords used to characterize a field of interest – the value of **Tf$_{jk}$** is accordingly large.

Thus, **Tf$_{jk}$** has the common behavior of a *Relevance* function. It indeed plays this role, when **TfIdf** is used to rank documents, with given characteristic keywords.

## 5.2   Idf for Unexpectedness

The *Inverse Document Frequency* **Idf** of keyword **k** is usually given by:

$$\mathbf{Idf_k = \log (N / df_k)} \qquad (5\text{-}2)$$

where **N** is the total number of documents in the sub-space under consideration and **df$_k$** is the number of documents containing keyword **k**.

**Idf$_k$** is inversely proportional to the number of documents of interest. The value of **Idf$_k$** increases when there are fewer documents containing the keyword **k**. This rewards document rarity. With this respect, the logarithm serves as a scaling factor, which does not change the **Idf$_k$** meaning.

Thus, **Idf$_k$** plays the role of an *Unexpectedness* function within the **TfIdf** criterion.

## 5.3   Interesting Uses of TfIdf

**TfIdf** has the form of a multiplicative composition, of the type of eq. (2-2):

$$\mathbf{TfIdf\ =\ Tf_{jk}\ *\ Idf_k} \qquad (5\text{-}3)$$

The **TfIdf** criterion is bipolar since it has exactly two functions pulling frequencies in opposite senses.

The common use of **TfIdf** is that of a ranking criterion for documents of interest, given keywords of relevance.

But, **TfIdf** can also be used the other way round. If **TfIdf** is indeed a valid criterion for interesting documents, given those documents it can be applied to find keywords which characterize the chosen field of interest. This is illustrated in the next case study.

## 5.4   Case Study: Finding Keywords of Relevance

This case study refers to software reuse of models and code found in the Web. In particular we were interested in cases that combine two design patterns, such as Observer and Mediator in the same code.

Search with the Google search engine started with keywords fetched from a target file containing both general words and keywords associated to specific design patterns.

Table 1: TfIdf values for Selected Keywords.

| Keyword | TfIdf |
|---|---|
| patterns | 14.14 |
| mediator | 12.47 |
| concatenate | 10.48 |
| observer | 6.27 |

The purpose of the **TfIdf** calculation was to find new keywords of relevance to the chosen field of software pattern reuse. In this context a document is the title and summary of each item in the search results. The candidate keywords were all words appearing in the search results.

Sorting the words by their **TfIdf** value produced the outcomes as seen in Table 1 – for a search of "observer mediator design patterns".

Although the number of appearances of the word "observer" is much higher than the word "concatenate", **TfIdf** values reverse their order and actually highlight new words of relevance, such as "concatenate". These new keywords were incorporated in the target file, for posterior use.

## 6   DISCUSSION AND RELATED WORK

The most striking feature of the vast literature on interestingness that we wish to convey in this very short literature review is the diversity of concepts and formulas: a broadly accepted framework is still lacking.

*Deviation* – in a statistical sense – has been used to characterize interestingness for automatic knowledge discovery in relatively early works – see e.g. (Piatetsky-Shapiro and Matheus, 1994 [9]).

A good source of references is the survey by Tuzhilin (Tuzhilin, 2000 [10]) in the Handbook of Data Mining and Knowledge Discovery (Klosgen and Zytkow, 2002 [4]), and references therein, e.g. (Padmanabhan and Tuzhilin, 1999 [8]).

Tuzhilin refers to three subjective measures of interestingness and ways to integrate them. Unexpectedness is explicitly mentioned. Another measure, *interestingness templates* reflects to a certain extent a form of relevance. The third one, *actionality* – another name for usefulness – is orthogonal to this paper's claim.

Arimura in reference (Arimura, 2000 [1]) uses the notion of *important* instead of interesting keywords. Among other techniques it mentions Shannon's *entropy* as a measure to discover important patterns. Information entropy is found in a variety of works – see e.g. (Li, 2006 [5]) and references therein.

A more recent survey of interestingness measures for knowledge discovery is found in (McGarry, 2005 [6]), from which one can infer that heterogeneity still characterizes the discipline.

## 6.1   A Unifying Paradigm

Against the background of so much diversity of content and form, this work offers a unifying conceptual paradigm of interestingness.

The unifying paradigm has a concise formal framework. Interestingness is the mathematical composition of exactly two functions: one standing for *relevance* to a chosen field of interest; the other for the *unexpectedness* that calls our attention to specific newly acquired knowledge.

This formal framework is clear enough to enable judgment relative to our intuitive notions and to the prevalent trends of research in the area.

On the other hand, the formal framework of this paradigm is not too restrictive. It encompasses functions currently used in practice and serves to stimulate findings of novel functional forms that fit the paradigm, as discussed next.

## 6.2   Varieties of Bipolar Composition

To esablish the viability of the unifying paradigm and its formal framework, it is important to show that many functional forms obey the paradigm.

We offered three examples, with their respective case studies: a multiplicative Match*Mismatch bipolar expression; a non-commutative pair of computational functions with application of Low and High thresholds; the well-known Tf*Idf criterion, which is also multiplicative. Each of them is bipolar in the sense that they involve exactly two functions, standing for Relevance and Unexpectedness.

As a counter-example to stimulate further research, we mention algebraic similarity vectors. Any such vector clearly represents the Relevance side of a possible bipolar expression. By itself it is not enough to express interestingness as required by the unifying paradigm.

One could define a kind of dissimilarity vector to represent unexpectedness, which is a subject for future investigation.

## 6.3   Main Contribution

The main contribution of this work is a unifying conceptual paradigm of interestingness for knowledge discovery:

-Mathematical composition of exactly two functions pulling in opposite directions, viz. Relevance and Unexpectedness.

## REFERENCES

[1] Arimura, H. et al., "Text Data Mining: Discovery of Important Keywords in the Cyberspace", Int. Conf. on Digital Libraries: Research and Practice, pp. 220-226, Kyoto, Japan (2000).
[2] Fujimura, K., et al., "The EigenRumor Algorithm for Ranking Blogs", in *WWW2005* Conf., Chiba, Japan (May 2005).
[3] Glance, N. S. et al., "BlogPulse: Automated Trend Discovery for Weblogs", in *WWW2004*, New York, NY USA, (May 2004).
[4] Klosgen, W. and Zytkow, J.M. (eds.), "Handbook of Data Mining and Knowledge Discovery", Oxford University Press, Oxford, Japan (2002).
[5] Li, G, et al., "Shortening retrieval sequences in browsing-based component retrieval using information entropy", J. Systems and Software, 79, pp. 216-230, (2006).
[6] McGarry, K., "A survey of interestingness measures for knowledge discovery", Knowledge Engineering Review J., 20 (1), 39-61 (2005).
[7] Moukas, A., "Amalthea: Information Discovery and Filtering using a Multiagent Evolving Ecosystem", Applied Artificial Intelligence, Vol. 11, Number 5, 1 pp. 437-457 (1997).
[8] Padmanabhan, B. and Tuzhilin, A., "Unexpectedness as a measure of interestingness in knowledge discovery", *Decision Support Sys.*, Vol. 27, (3) (1999).
[9] Piatetsky-Shapiro, G. and Matheus, C.J., "The Interestingness of Deviations", KDD-94, AAAI-94 Knowledge Discovery in Databases Workshop (1994).
[10] Tuzhilin, A., "Usefulness, Novelty, and Integration of Interestingness Measures", chapter 19.2.2 in reference [4], pp. 496-508 (2002).
[11] Yuwono, B. et al., "A WWW Resource Discovery System", in 4[th] Int. WWW Conf., Boston, MA., USA (December 1995).